# Compact Photonic Fibre-based Deformation Sensor Fabricated by Two-Photon Polymerization

**Vladimir Osipov[a]*, Ayman El-Tamer[b], Ulf Hinze[b,c], Alexander Dulebo[d] and David J. Webb[a]**

[a]Aston Institute of Photonic Technologies, Aston University, B4 7ET, Birmingham, UK

[b]Laser nanoFab GmbH, Hollerithallee 17, 30419 Hannover, Germany

[c]Leibniz Universität Hannover, Welfengarten 1, 30167 Hannover, Germany

[d]Bruker Nano GmbH, Am Studio 2D, 12489 Berlin, Germany

**Abstract.** We have demonstrated that compact deformation sensors with heights and widths of about 100 μm can be fabricated by two-photon polymerization, using commercially available optical ferrules with embedded 125 μm-diameter optical fibers as the basic platform and the commercial photopolymers OrmoComp and FemtoBond as the fabrication materials. For both fabricated and tested sensor types, forces as small as 10 nN could be resolved, and a reasonably linear relationship between force and deformation, expressed as membrane displacement, was observed over the measured force range of 200–3600 nN.

*Vladimir Osipov, v.osipov@aston.ac.uk; v.osipov.research@gmail.com

## 1 Introduction.

The Two-Photon Polymerization (2PP) laser fabrication technique Ref.1 allows for direct 3D fabrication of designed microstructures with micron or sub-micron scale precision: this is an exceptionally flexible approach with results that are not achievable by established additive manufacturing techniques and which leads to a range of novel device design possibilities. For example, 2PP has been used to fabricate a set of compact optical components (lenses, waveguides, gratings, as well as hybrid diffractive–refractive micro-optical elements, Ref 2) and can be used to create the micro-fluidic structure of a device incorporating precision mechanical components such as locks and shutters Ref.3. It has previously been used for the fabrication of micro-optics for biomedical applications Ref.4,5 or creation of diffractive optical elements producing designed 3D laser intensity distributions for drug delivery Ref.6.

Later this 2PP technology was used for the realization of compact pressure Ref.7 and acoustic Ref.8 sensors on optical fibir end-facets, where the sensing element was a thin 2PP-fab-ricated polymer membrane. Its displacement under static pressure or acoustic pressure was measured using a laser beam supplied from the fibre core and reflected from the membrane. The main purpose of this work is to demonstrate the capability of 2PP laser technology to enable the design and direct fabrication of compact deformation sensors on optical fibre end-facets, as a platform, and fabrication of associated polymer membranes with thickness down to 3 μm. To hold the membranes, we propose to use fibre-in-ferrule patches – tools developed by the photonics industry – as they nicely complement commercial optical fibre patch cords and can also be used separately.

## 2 Methods

### *2.1 Design*

We used COMSOL Multiphysics® 5.4 software to create our micro-sensor design. This simulation platform provides fully coupled multi-physics modelling capabilities. It includes all the steps in the modelling workflow — from defining geometries, material properties, and the physics that describe specific phenomena to performing computations and evaluating the results. 3D design samples, developed in COMSOL, can be easily exported in stl-file format, that can be used for direct 3D fabrication on optical fibre end-facet by means of a 2PP system with an appropriate optical fibre holder.

Using the COMSOL Multiphysics with Wave Optics Module, preliminary designs were completed of compact deformation micro-sensors oriented to 2PP-fabrication on the optical fibre end-facet, consisting of a micro-lens, optical cavity and 3μm-thickness membrane - see Fig.1 (a - basic cylinder-type, b – suspended 3ray star-type.

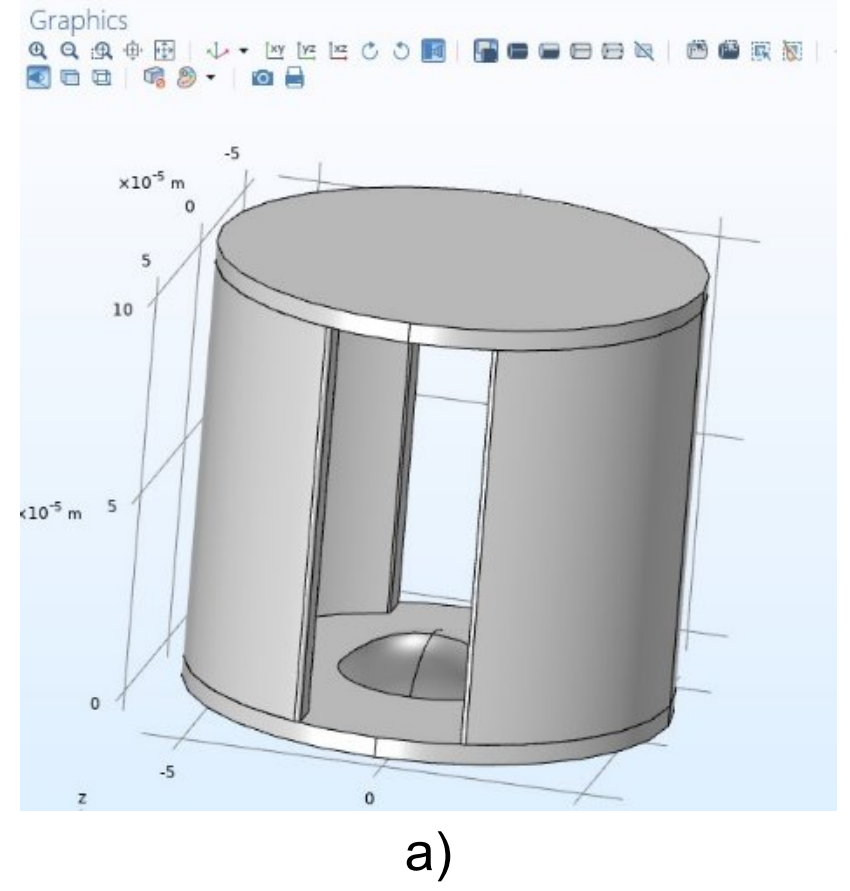

a)

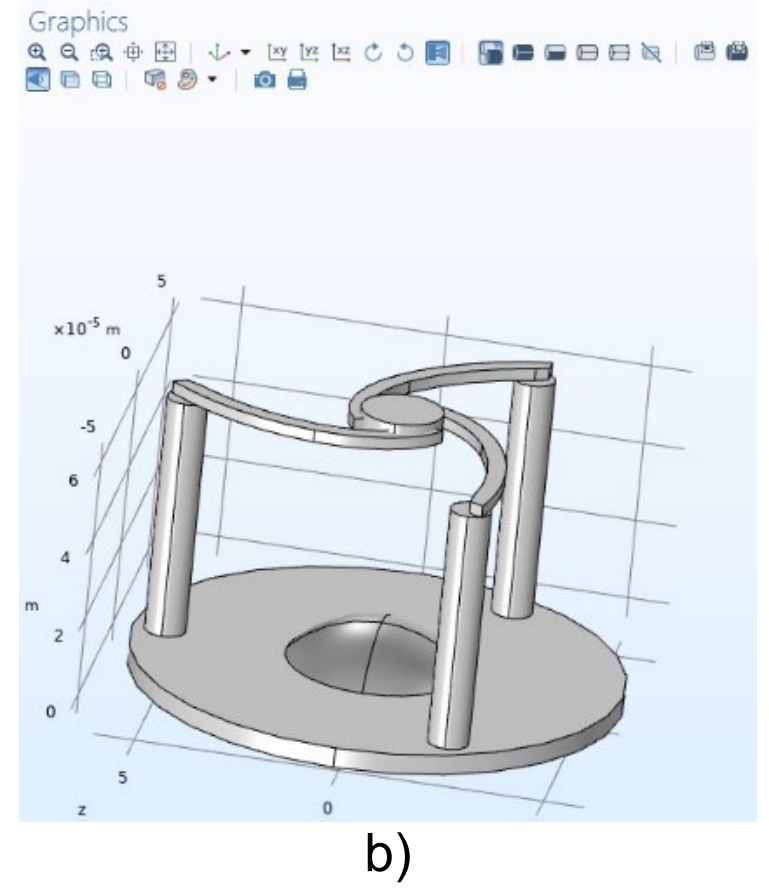

b)

Fig.1. COMSOL script files of the deformation micro-sensors based on: micro-lenses, membranes and optical cavities, designed for 2PP fabrication on optical fibre –end-facet (diameter 120 μm). a) – basic cylinder-type, b) – suspended star-type.

### *2.2. 2PP setup*

Our experimental 2PP setup was designed in cooperation with Laser nanoFab (Germany) and is based on a Chromacity Spark 1040 laser with SHG module (wavelength 520 nm, re-petition rate 100 MHz, pulse length 120 fs, average power on sample up to 100 mW), AOM for laser power application control, 3D XYZ nano-stages with travel range 5 mm and micromirror-based laser scanner for rapid fabrication on fibre top-sized areas. Our setup was adapted to allow the designed fundamental components to be fabricated on both planar glass substrates and optical fibre tops. For fabrication on fibre top there was realised a special fibre holder (Fig.2), which allows 3D precise adjustment of the fibre top relative to the focussing micro-objective position.

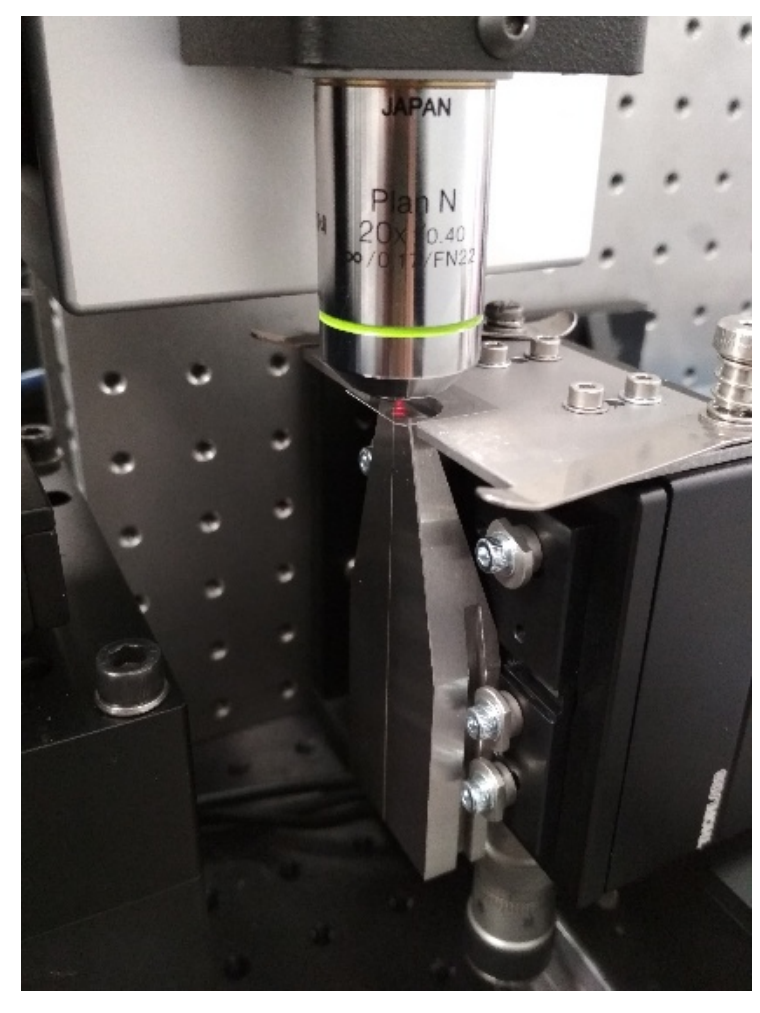

Fig.2. The optical fibre holder for flexible 2PP fabrication on glass substrates or 125 µm diameter optical fibre end-facets.

**Fabrication parameters:** Scannermode, Speed: 2mm/s, Power: 6 mW for FemtoBond 4C and 7 mW for OrmoComp, Slicing: 500 nm, Hatching: 200 nm.

## 3 Results

Images of fabricated micro-sensors of cylinder and suspended star designs with 3 µm thickness membrane on glass substrates and on an optical fibre end-facet are shown in Fig.3.

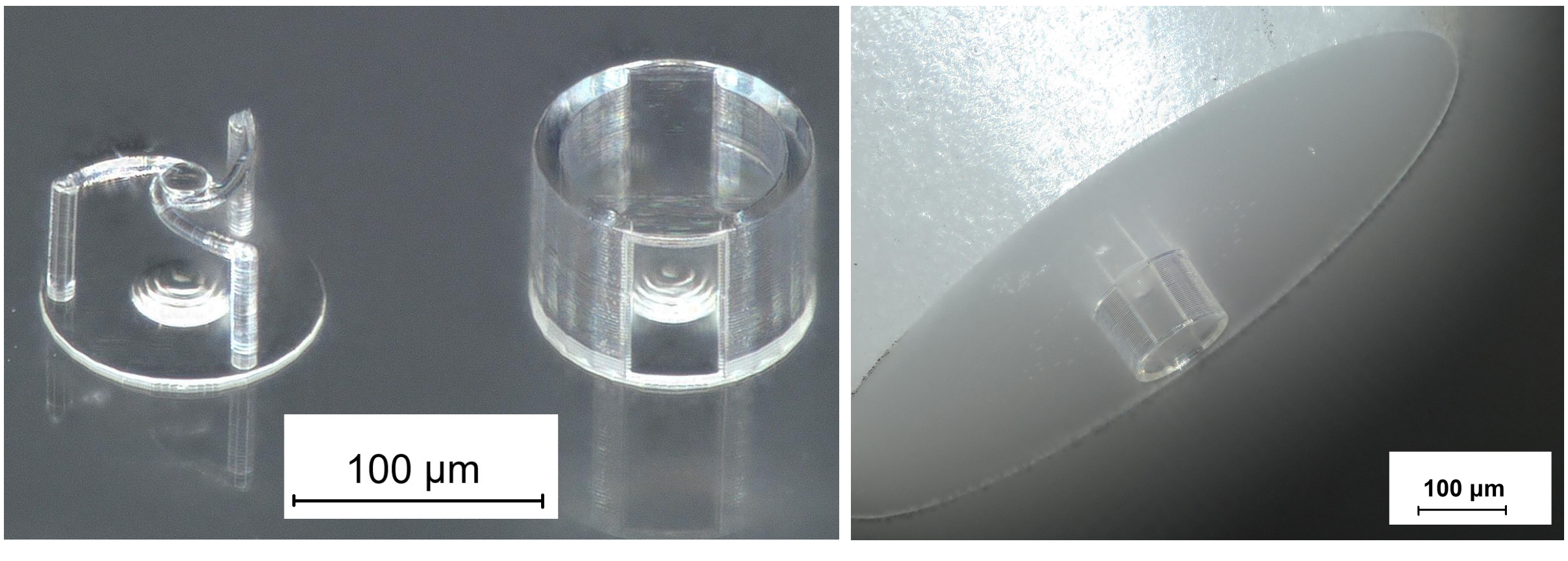

a) b

Fig.3. Microscope images of the fabricated micro-sensors on (a) glass substrate and (b) 125 µm-diameter optical fibre end-facet embedded in a 1.25mm ferrule. Markers in the lower right corner are equal to 100 µm.

## 4 Atomic Force Microscopy (AFM) characterisation of the fabricated Micro-Sensors

We carried out an AFM study to characterise the deformation sensitivity of the micro-sensors fabricated on glass substrates.

**Equipment Used:** Bruker Dimension Icon Atomic Force Microscope with inspectable area 180x150x40 mm$^3$, noise < 30 pm RMS, 90x90x10 μm$^3$ scan size. Elasticity measurements were done by pressing on a central part of the construct using the AFM tip. The SEM views of the AFM probe are shown in Fig.4a,b. To apply various loading forces, two different AFM probes (both from Bruker) with different spring constants were used: RFESPA-75 (1.21 N/m), and RTEPSA-300-125 (40.6 N/m). The approximate position of the tip-sample contact is shown using a red cross on the optical image Fig.4c.

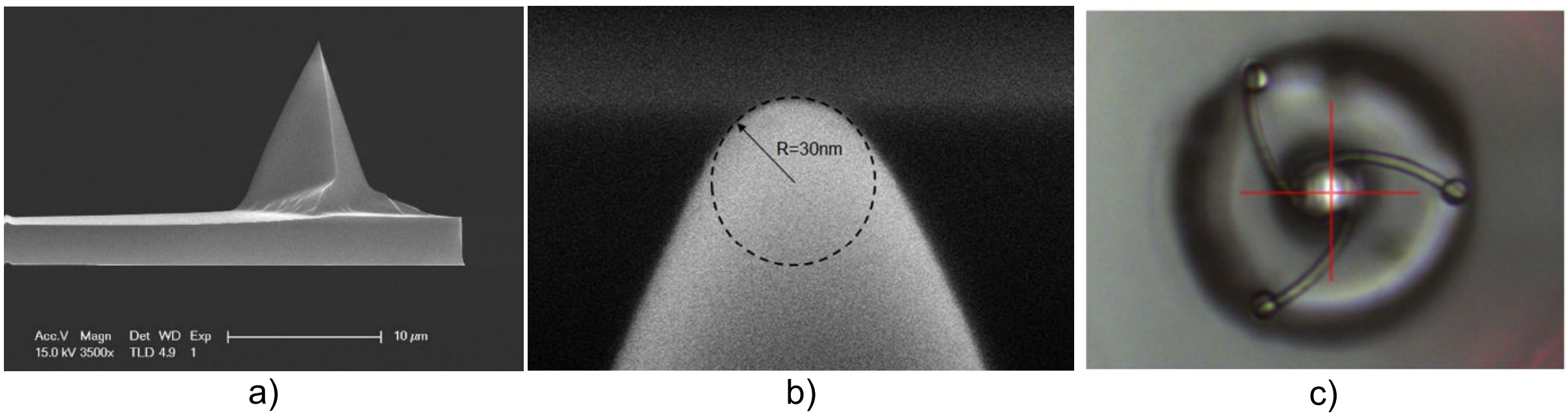

a) b) c)

Fig.4. a) SEM view of AFM probe RTEPSA-300-125 (40.6 N/m); b) tip of the probe; c) approximate position of the tip-sample contact is shown by a red cross on image of the sensor (sensor diameter is 120 μm).

Images of glass substrates with 2PP fabricated micro-sensors, using FemtoBond and Ormo-Comp, are shown in Fig. 5 and 6, respectively.

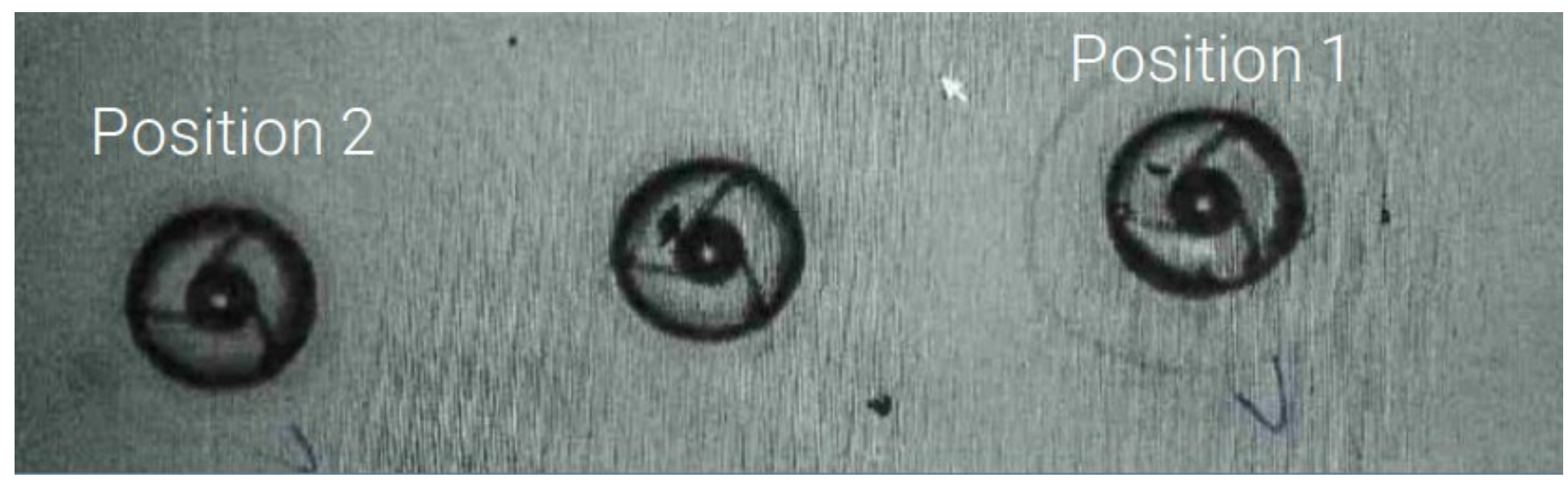

Fig.5. FemtoBond photopolymer-based 2PP fabricated sensors. Position 1 and 2: tested 3-star type sensors, outer diameter is 120 μm.

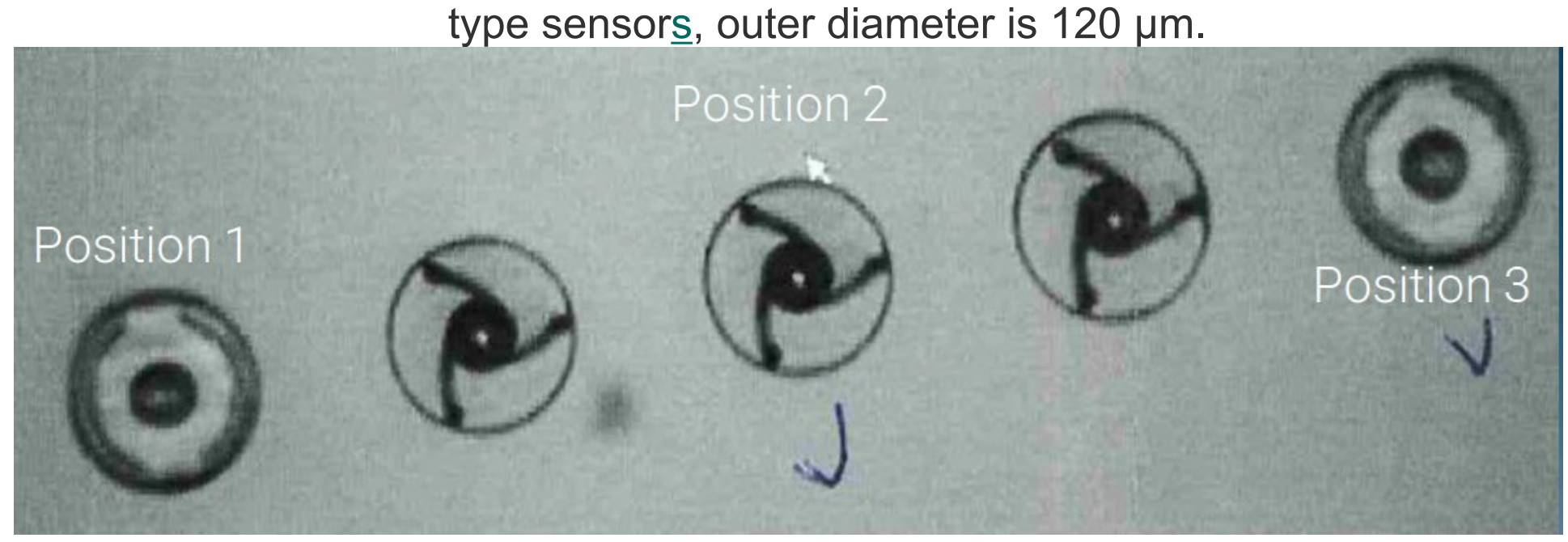

Fig.6. OrmoComp photopolymer based 2PP fabricated samples. Position 1 and 3: Basic type sensor; Position 2: tested 3-star type sensor, outer diameter is 120 μm.

Data in the form of force-distance curves were collected. The curves for FemtoBond-based micro-sensor Position 2 (3-star type) were processed, and sample stiffness and sample deformation were measured and presented in Fig.7.

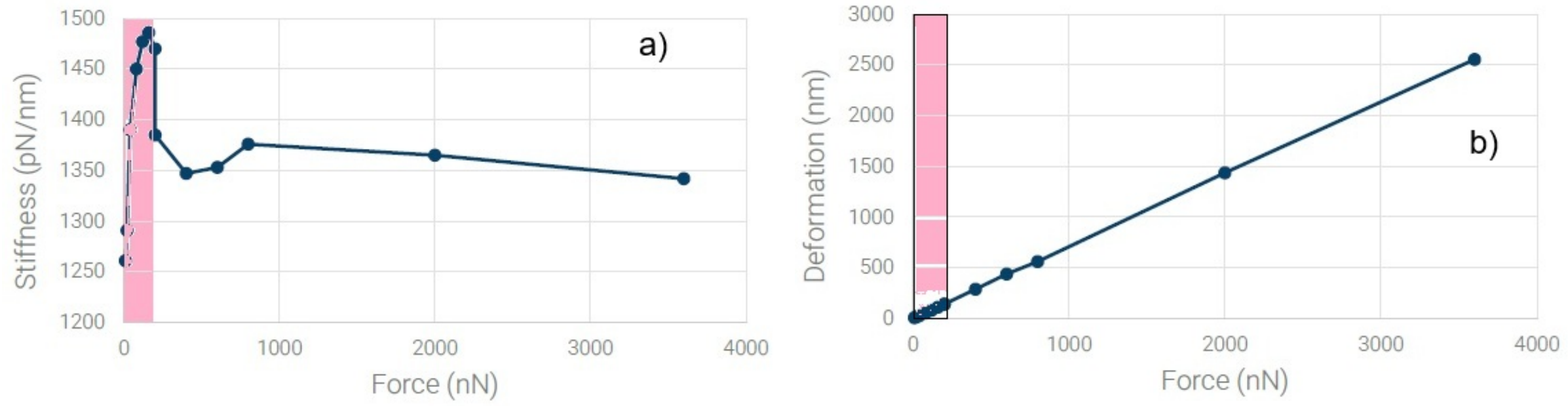

Fig.7.The 3-star type micro-sensor, based on FemtoBond – Position 2 in Fig.5: a) stiffness and b) deformation depending on the force, applied to probe; RFESPA-75 (rose colour marked) and RTEPSA-300-125 probes were used with probes changed around 200 nN force magnitude.

Data in the form of deformation-stiffness curves vs applied force were analysed for different types of OrmoComp-based sensors (basic & 3-star types), as shown in Fig.8.

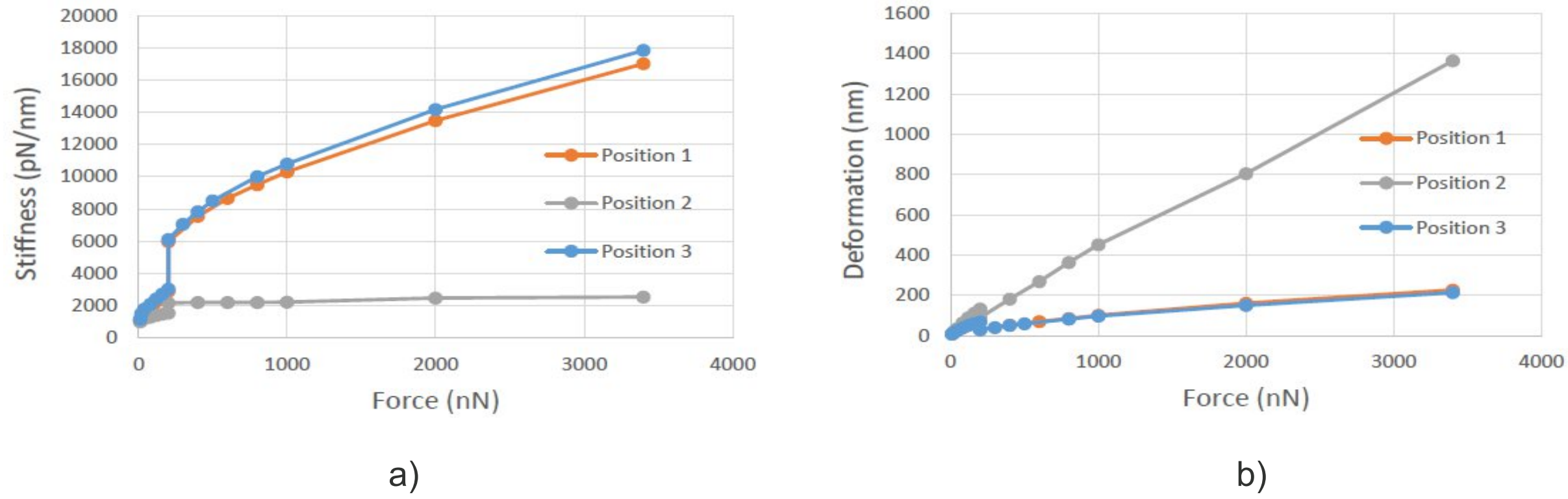

a) b)

Fig.8. Stiffness (a) and deformation (b) dependences on the applied force curves for basic type (Position 1 and 3 in Fig. 6) and 3-star type (Position 2 in Fig. 6) micro-sensors based on OrmoComp.

It was shown for both types of fabricated sensors that forces as small as 10 nN can be resolved and good proportionality of deformation (that is distance movement) vs force is observed in the measured range 200-3600 nN. Good proportionality of deformation vs force was also observed in the range 10-100nN by the more sensitive probe. The data in the range 100-200 nm is affected by switching between the probes with different sensitivity. The achieved sensitivity of 2PP fabricated micro-sensors is good enough for applications that require precise deformation measurements Ref 9.

## Conclusions

It was demonstrated that 2PP fabrication of compact deformation sensors with heights and width about 100 µm is possible, using as a basic platform commercially available optical ferrules with embedded 125 µm-diameter optical fibres and commercial photopolymers OrmoComp and FemtoBond. It was shown, that for both types of designed, fabricated and tested sensors (basic and 3-ray star types), forces as small as 10 nN can be resolved and good proportionality of deformation vs force (that is membrane displacement) is observed in the measured force range 200-3600 nN.

## Disclosures

The authors declare that there are no financial interests or other potential conflicts of interest that could have influenced the objectivity of this research or the writing of this paper.

## Code and Data Availability

All datasets used in this work are publicly available through the corresponding references listed. In the meantime, specific inquiries can be directed to the corresponding author.

## Acknowledgements

We would like to gratefully acknowledge the LEVERHULME TRUST for support of this work, Research Project Grant RPG-2022-334. Ayman El-Tamer and Ulf Hinze acknowledge funding from the European Union (ERC, Laser-Tissue-Perfuse, 101054009).

**Vladimir Osipov** has been a researcher with Stepanov Institute of Physics (SIP), Belarus since 1976, Marie-Curie Fellow with the Laser Zentrum Hannover (LZH), Germany during 2010-2012 and Research Associate with Aston University, UK since 2017. He received his PhD in quantum electronics from SIP (1997). He has been awarded 9 patents and has more than 75 research publications. His interests include two-photon polymerization, micro-optics/micro-devices design and laser fabrication.

**Alexander Dulebo** is an application engineer with Bruker Nano, Inc., Germany since 2012. He received his PhD degree in biophysics from University of South Bohemia in České Budějovice, Czech Republic (2010). He has worked on various tasks, exploring life and materials at molecular, cellular and microscopic levels.

**David J. Webb** joined Aston University as Reader in Photonics in 2001, was promoted to Professor in 2012 and awarded a 50th Anniversary Chair in 2016. Prior to joining Aston University he had spent 10 years as a Lecturer, then Senior Lecturer in the Physics Department at the University of Kent at Canterbury. He received BA in Physics (1982) at the University of Oxford and the PhD in Physics (1988) at the University of Kent. His professional work has primarily focused on solving complex object detection and tracking problems across a variety of domains, including overhead imagery, robotics, and real-time video analysis systems.

**Ayman El-Tamer** received his Diploma in Physics from Leibniz Universität Hannover, Germany, in 2013. After working as a research scientist at Laser Zentrum Hannover e.V., he joined Laser nanoFab GmbH in 2020, where he serves as shareholder and Head of Micro- and Nanotechnology. His research focuses on two-photon polymerization (2PP), micro- and nanofabrication, and their biomedical applications. He is responsible for 2PP process and system development and has contributed to numerous EU-funded research projects, scientific publications, book chapters, and patents.

**Ulf Hinze** is co-founder and managing director of Laser nanoFab GmbH, Germany, since 2017. He received his Diploma in Physics from the University of Hannover in 1996 and his Dr. rer. nat. in physics from the Institute of Quantum Optics, University of Hannover, in 2000. From 2001 to 2017, he held several group-leader positions at Laser Zentrum Hannover e.V. (LZH), including Laser Metrology, EUV and X-ray, and Nanolithography. Since 2019, he has been a member of the Cluster of Excellence QuantumFrontiers at Leibniz Universität Hannover, where he also contributes to research and teaching. His work focuses on laser micro- and nanofabrication, two-photon polymerization, micro-optics, optical metrology, and EUV/X-ray technologies.